\documentstyle [12pt] {article}
\begin{document}
\def\be{\begin{eqnarray}}
\def\en{\end{eqnarray}}
\def\non{\nonumber}
\def\la{\langle}
\def\ra{\rangle}
\def\ep{\varepsilon}
\def\K{{K^{(*)}}}
\def\D{{D^{(*)}}}
\def\r{{\tau(B^-)\over\tau(\bar{B}^0)}}
\def\pr{{\sl Phys. Rev.}~}
\def\prl{{\sl Phys. Rev. Lett.}~}
\def\pl{{\sl Phys. Lett.}~}
\def\np{{\sl Nucl. Phys.}~}
\def\zp{{\sl Z. Phys.}~}

\font\el=cmbx10 scaled \magstep2
{\obeylines
\hfill IP-ASTP-21-94
\hfill November, 1994}

\vskip 1.5 cm

\centerline{\large\bf Extraction of $a_1$ and $a_2$ from $B\to \psi K(K^*),~
D(D^*)\pi(\rho)$ Decays}
\medskip
\bigskip
\medskip
\centerline{\bf Hai-Yang Cheng and B. Tseng}
\medskip
\centerline{ Institute of Physics, Academia Sinica}
\centerline{Taipei, Taiwan 11529, Republic of China}
\bigskip
\bigskip
\bigskip
\centerline{\bf Abstract}
\bigskip
{\small   Based on the factorization approach,
we show that the CLEO data for the ratio $\Gamma(B\to\psi K^*)/\Gamma(B\to
\psi K)$ and the CDF measurement of the
fraction of longitudinal polarization in $B\to\psi K^*$
can be accounted for by the heavy-flavor-symmetry approach
for heavy-light form factors provided that the form factor $F_0$ behaves
as a constant, while the $q^2$ dependence is of the monopole form for $F_1,~
A_0,~A_1$, and of the dipole behavior for $A_2$ and $V$. This $q^2$
extrapolation
for form factors is further supported by $B\to K^*\gamma$ data and by a recent
QCD-sum-rule analysis. We then apply this method to $B\to D(D^*)\pi(\rho)$
decays to extract the parameters $a_1$ and $a_2$. It is found that
$a_1(B\to D^{(*)}\pi(\rho))=1.01\pm 0.06$ and $a_2(B\to
D^{(*)}\pi(\rho))=0.23\pm 0.06\,$. Our result $a_2/a_1=0.22\pm 0.06$ thus
significantly improves the previous analysis that leads to $a_2/a_1=0.23\pm
0.11$. We argue that, contrary to what anticipated from the leading $1/N_c$
expansion, the sign of $a_2(B\to\psi\K)$ should be positive and
$a_2(B\to\psi\K)~{\large ^>_\sim}~ a_2(B\to D^{(*)}\pi(\rho))$.
}

\pagebreak

   {\it\bf 1. Introduction}~~~
 The fact that the $D^+$ meson has a longer lifetime than the $D^0$ is already
manifest at the exclusive two-body decay level: The number of two-body $D^+$
decay modes is about two times less than that of the $D^0$ and
there exists a large destructive interference in the Cabibbo-allowed decays
$D^+\to \bar{K}^0\pi^+,~\bar{K}^0\rho^+,~\bar{K}^{*0}\pi^+$, which is also
known as the Pauli interference at the inclusive level.
The recent CLEO data on ${B}\to D\pi,~D\rho,~D^*\pi,~D^*\rho$ decays exhibit
a rather unexpected result [1]: The
interference between the two different amplitudes contributing to exclusive
two-body $B^-$ decays are evidently constructive, contrary to the charmed
meson case. This feature is quite stunning since the rule of retaining only
the leading terms in the $1/N_c$ expansion ($N_c$ being the number of color
degrees of freedom) [2], which is empirically operative in charm decays,
fails in $B\to D^{(*)}\pi(\rho)$ decays. Quantitatively, the ratio of
the parameters $a_1$ and $a_2$ corresponding to the
external and internal $W$-emission amplitudes
is found to be positive with the magnitude $a_2/a_1=0.23\pm 0.04\pm 0.04\pm
0.10$ [1].

    Under the factorization assumption, the spectator meson decay amplitude
is characterized by the parameters $a_1$ and $a_2$ [3] which are related to
the Wilson coefficients $c_1$ and $c_2$ by
\be
a_1=\,c_1+\xi_1c_2,~~~~a_2=\,c_2+\xi_2 c_1.
\en
Naive factorization implies that $a_1$ and $a_2$ are
universal, namely they are channel independent in $D$ or $B$ decays.
Recently, we have shown from the analysis of experimental data that $a_2$ is
not universal at least in charm decays [4]. We found $a_2\sim -0.51,~-(0.66
\sim 0.78),~-(0.85\sim 0.91)$ respectively for $D\to\bar{K}\pi,~\bar{K}^*\pi$
and $\bar{K}^*\rho$ decays [4]. Physically, this can be understood
as follows. Writing
\be
\xi_1=\,{1\over N_c}+{r_1\over 2},~~~~~\xi_2=\,{1\over N_c}+{r_2\over 2},
\en
where $r_{1,2}$ denote the contributions of color octet currents arising from
the Fierz transformation relative to the factorizable ones [5],
\footnote{Soft gluon contributions are of course nonfactorizable. The fact
that $r_{1,2}$ are channel dependent reminds us of that they are originally
nonfactorizable.}
it is natural to
expect that the nonperturbative effect is such that $|r_2(D\to VV)|>|r_2
(D\to VP)|>|r_2(D\to PP)|$ ($V$: vector meson, $P$: pseudoscalar meson)
since soft-gluon effects become more important when final-state particles
move slower, allowing more time for significant final-state interactions.
Numerically, it follows from Eq.(2) that $r_2\sim -0.67,~-(0.9\sim 1.1),
{}~-(1.2\sim 1.3)$ respectively in $D\to\bar{K}\pi,~\bar{K}^*\pi$ and
$\bar{K}^*\rho$
decays [4], in accordance with the theoretical expectation. It is clear that
the leading $1/N_c$ expansion works most successfully for $D\to\bar{K}\pi$
as the subleading $1/N_c$ factorizable contribution is almost compensated
by the soft-gluon effect so that $\xi_2(D\to\bar{K}\pi)\approx 0$.

    Now come back to the $B$ meson case. {\it A priori} there is no reason
to expect that $a_2$ extracted from $B\to\psi K^{(*)}$ should be the same as
that in $B\to D^{(*)}\pi(\rho)$ decays. Recall that the c.m. momentum in
$B\to D\pi$ is 2306 MeV, while it is only 1682 MeV in $B\to\psi K$ decay.
Hence a direct determination of $a_2$ from $B\to D^{(*)}\pi(\rho)$ data is
desirable and it is expected that $|r_2(D\to PP)|>|r_2(B\to \psi K)|~{\large
^{>}_{\sim}}~|r_2(B\to
D\pi)|$. However, it is easily seen from Tables XX and XXI of Ref.[1] that,
based on the modified Bauer-Stech-Wirbel model [6],
an individual fit of $a_2/a_1$ to the CLEO data of $B\to \D\pi(\rho)$
gives rise to
\be
{a_2\over a_1}=\cases{0.31\pm 0.12\,, & $B\to D\pi$; \cr  0.44\pm 0.23\,, &
$B\to D\rho$; \cr   0.32\pm 0.13\,, & $B\to D^*\pi$; \cr 0.68\pm 0.26\,, &
$B\to D^*\rho$. \cr}
\en
Since $r_{1,2}$ are not supposed to vary significantly from
$B\to D\pi$ to $D^*\rho$ decays, a sizeable discrepancy among some of
the values of $a_2/a_1$ shown in (3) determined from various $B$ decay
modes is certainly unexpected. Recall that the aforementioned value $a_2/a_1=
0.23\pm 0.11$ cited in the CLEO paper [1], which is substantially different
from the individual fits given in (3), is obtained by a global least
squares fit to the CLEO data. Of course, it
is more complicated to extract $a_2$ from $B\to D^{(*)}\pi(\rho)$ decays than
from $B\to \psi K^{(*)}$ since the former involve final-state interactions and
$W$-exchange contributions. Nevertheless, even in the absence of the above
two effects, a better improvement on the previous theoretical calculations
for $B\to D^{(*)}\pi(\rho)$ is called for.

  It is tempted to argue that the above-mentioned difficulty for
extracting $a_2$ is circumvented in the case of $B\to \psi K^{(*)}$ decays
as they are free of final-state strong interactions and nonspectator effects.
Indeed, an individual fit of $a_2$ to
the CLEO data of $B^-\to\psi K^-,~\psi K^{*-}$, $\bar{B}^0\to\psi\bar{K}^0,
{}~\psi\bar{K}^{*0}$ yields $|a_2|=0.25\pm 0.02,~0.25\pm 0.04,~0.20\pm 0.03,~
0.24\pm 0.03$, respectively (see
Tables IX and XX of Ref.[1]). However, it was pointed out recently that there
are two experimental data for $B\to\psi K^{(*)}$ which cannot be accounted for
simultaneously by all commonly used models [7,8]. That is, all the
known models in the literature fail to reproduce the data of either the
production ratio $R=\Gamma(B\to\psi K^*)/\Gamma(B\to\psi K)$ or the fraction
of longitudinal polarization $\Gamma_L/\Gamma$ in $B\to \psi K^*$ or both.
This casts doubt on the estimate of $a_2$ from $B\to\psi K^{(*)}$ and even
on the validity of the factorization approach.

  In this paper, we will explore the possibility of accounting for the data of
$B\to \psi K^{(*)}$ without giving up the factorization hypothesis. We found
that the data of $R$ measured by CLEO and $\Gamma_L/\Gamma$ by
CDF for $B\to\psi K^{(*)}$ decays can be accommodated by the heavy flavor
symetry approach for heavy-light form factors with an appropriate choice
for their $q^2$ dependence. This method for heavy-light form factors is then
applied to
$B\to D^{(*)}\pi(\rho)$ decays and compared with experiment. We shall see that
$a_2/a_1$ or $a_2$ extracted in this way is improved quite significantly. A
discussion on its implication is then presented.

\vskip 0.3cm
  {\it\bf 2. Implications from $B\to\psi K^{(*)}$ decays}~~~The experimental
data of interest for $B\to\psi\K$ are the vector to pseudoscalar production
ratio [1]
\be
R=\,{{\cal B}(B\to\psi K^*)\over {\cal B}(B\to \psi K)}=\,1.71\pm 0.34\,,
\en
and the fraction of longitudinal polarization in $B\to \psi K^*$
\be
\left({\Gamma_L\over\Gamma}\right)_{B\to\psi K^*}=\cases{ 0.97\pm 0.16\pm
0.15, & ARGUS~[9]; \cr 0.80\pm 0.08\pm
0.05, & CLEO~[1]; \cr 0.66\pm 0.10^{+0.08}_{-0.10},& CDF~[10].  \cr}
\en
Based on factorization, it was shown recently that currently used $B\to
K^{(*)}$ form factors fail to explain the data of $R$ and $\Gamma_L/\Gamma$
simultaneously [7,8]. This might be attributed either to a failure of
the factorization method (see e.g. Ref.[11]) or to a wrong choice of form
factors or both.

In what follows we would like to investigate if it is possible to ``derive"
a set of $B\to K^{(*)}$ form factors to account for the $B\to\psi K^{(*)}$
data within the factorization framework. Since the existing models
lead to form factors excluded by data [7,8],
we prefer to follow Ref.[12]
to relate the $B\to K^{(*)}$ and $D\to K^{(*)}$ form factors at the same
heavy quark velocity $v$ via model-independent heavy flavor symmetry, as
elaborated on in Ref.[7]. So long as the momentum of the light meson $p_\K$
does not scale with $m_{c,b}$ or $v\cdot p_\K<<m_{c,b}$, one has the
relations [12]
\footnote{Empirically, we found that the heavy-light form-factor relations (6)
using the heavy quark masses $m_b$ and $m_c$ work better for $B\to\psi\K$
decays than that using the meson masses $m_B$ and $m_D$.}
\be
F_1^{BK}(q_B^2) &=& {C_{bc}\over 2\sqrt{m_bm_c}}\left\{ (m_b+m_c)F_1^{DK}(q_D
^2)-(m_b-m_c){m_D^2-m_K^2\over q^2_D}\left[F_0^{DK}(q_D^2)-F_1^{DK}(q_D^2)
\right]\right\},   \non \\
V^{BK^*}(q_B^2) &=& C_{bc}\sqrt{m_c\over m_b}\,{m_B+m_{K^*}\over m_D+m_{K^*}}
V^{DK^*}(q_D^2),  \non  \\
A_1^{BK^*}(q_B^2) &=& C_{bc}\sqrt{m_b\over m_c}\,{m_D+m_{K^*}\over m_B+m_{K^*}}
A_1^{DK^*}(q_D^2),   \\
A_2^{BK^*}(q_B^2) &=& {1\over 2}C_{bc}\sqrt{m_c\over m_b}\Bigg\{\left(1+{m_c
\over m_b}\right){m_B+m_{K^*}\over m_D+m_{K^*}}A_2^{DK^*}(q_D^2)+\left(
1-{m_c\over m_b}\right){m_B+m_{K^*}\over q_D^2}   \non \\
&\times & \left[2m_{K^*}A_0^{DK^*}(q_D^2)-(m_D+m_{K^*})A_1^{DK^*}(q_D^2)
+(m_D-m_{K^*})A_2^{DK^*}(q_D^2)\right]\Bigg\}, \non
\en
where $C_{bc}=\left(\alpha_s(m_b)/\alpha_s(m_c)\right)^{-6/25}$, $q_B=m_bv
-q,~q_D=m_cv-q$, and we have followed Ref.[13] for the definition of form
factors.

    It is customary to make a monopole ansatz for all the form factors
$F_0,~F_1,~A_0,~A_1$, $A_2,~V$ [13]. However, a careful study based on the
scaling argument indicates that not all the form factors share the same $q^2$
behavior. A consideration of the heavy quark limit behavior of the form
factors leads to [14]
\be
F_0^{BK}(q^2)=\,F_1^{BK}(q^2)\left( 1-{q^2\over m_B^2-m_K^2}\right),
\en
and
\be
A_0^{BK^*}(q^2)=\,{m_B+m_{K^*}\over 2 m_{K^*}}A_1^{BK^*}(q^2)-{m_B^2
-m^2_{K^*}-q^2\over 2m_{K^*}(m_B+m_{K^*})}A_2^{BK^*}(q^2)
\en
to the leading order in the heavy quark limit.
Note that at $q^2=0$, (7) and (8) are precisely
the constraints [13]
\be
F_0^{BK}(0) &=& F_1^{BK}(0),   \non  \\
A_0^{BK^*}(0) &=& {m_B+m_{K^*}\over 2 m_{K^*}}A_1^{BK^*}(0)-{m_B-m_{K^*}
\over 2m_{K^*}}A_2^{BK^*}(0),
\en
necessary for avoiding unphysical poles on the r.h.s. of Eq.(6). It is
evident that the $q^2$ dependence of $F_1$ is different from that of $F_0$
by an additional pole factor. Several theoretical arguments and many QCD
sum rule calculations [15-18] indicate that $F_1(q^2)$ has a monopole behavior.
This in turn implies an approximately constant $F_0$.
\footnote{It was pointed out in Ref.[19] that if the single pole behavior
holds for both $F_0$ and $F_1$, one is led to $f_-^{B\pi}(0)=-0.21f_+
^{B\pi}(0)$, which is inconsistent with the heavy-quark-symmetry relation
$f_-^{B\pi}\approx -f_+^{B\pi}$. A nearly constant behavior of $F_0$ in $q^2$
is confirmed by a recent QCD-sum-rule calculation [17].}

    If $A_0$ and $A_1$ have the same $q^2$ dependence, Eq.(8) will lead to
an interesting $q^2$ behavior for $A_2$ [14]. Assuming
\be
A_0(q^2)=\,{A_0(0)\over\left(1-{q^2\over m^2}\right)^n},~~~~A_1(q^2)=\,{A_1
(0)\over\left(1-{q^2\over m^2}\right)^n},
\en
where the pole mass $m$ is the same for $A_0$ and $A_1$ in the heavy quark
limit, we see from Eq.(8) that, by neglecting $m_{K^*}$ relative to $m_B$,
the $q^2$ dependence of $A_2$ is different from that of $A_0$ and $A_1$
by an additional pole factor [14]:
\be
A_2(q^2)=\,{A_2(0)\over\left(1-{q^2\over m^2}\right)^{n+1}}.
\en
In practice, we will take $n=0,1$.

   In the following we will calculate $B\to\K$ form factors from Eq.(6) using
the experimental input on $D\to\K$ ones [20]:
\footnote{The average experimental values given in
the Particle Data Group [21] are
$F_1^{DK}(0)=0.75\pm 0.03\,,$ $V^{DK^*}(0)=1.1\pm 0.2,~~A_1^{DK^*}(0)=0.56
\pm 0.04,~~A_2^{DK^*}(0)=0.40\pm 0.08.$}
\be
F_1^{DK}(0) &=& 0.77\pm 0.04\,,~~~~V^{DK^*}(0)=1.12\pm 0.16\,,  \non \\
A_1^{DK^*}(0) &=& 0.61\pm 0.05\,,~~~~A_2^{DK^*}(0)=0.45\pm 0.09\,.
\en
As for the $q^2$ dependence, since $A_2$ has one more pole factor than $A_0$
and $A_1$, as just discussed, two possibilities of interest are:

   (i) a monopole form for $F_1,~A_2,~V$, and an approximately constant
for $F_0,~A_0,~A_1$. Recall that a slowly varying $A_1(q^2)$ with $q^2$ is
strongly advocated in Ref.[8].

   (ii) a monopole extrapolation for $F_1,~A_0,~A_1$, a dipole behavior for
$A_2,~V$, and an approximately constant for $F_0$. This is precisely the pole
behavior shown by a recent QCD sum rule analysis [18].
\footnote{The unusual $q^2$ behaviors for the form factors $A_1$ and $A_2$
obtained in existing QCD-sum-rule calculations [16] are no longer seen in
a recent similar work [18].}

\noindent In addition to the above two cases,
we also consider case (iii) in which all the form factors
are extrapolated in a monopole form, as employed in the Bauer-Stech-Wirbel
(BSW) model [13]. Given a set of extrapolation procedures, the
$D\to\K$ form factors are first extrapolated from $q^2=0$ to maximum $q_D^2$,
and then related to the $B\to\K$ form factors at zero recoil via Eq.(6).
Using $m_b=5$ GeV and $m_c=1.5$ GeV, the calculated $B\to
\K$ form factors at $q^2=0$ and $m_\psi^2$ are summarized in Table I.
For a comparsion, we have included in Table I two other model predictions: (1)
the BSW model (BSWI) [13] in which the $B\to\K$ form factors are first
calculated at $q^2=0$ and then extrapolated to finite $q^2$ using a monopole
behavior for all the form factors, and (2) the modified BSW model (BSWII)
[5], which is the same as BSWI except for a dipole $q^2$ dependence for form
factors $F_1,~A_0,~~A_2,~V$, inspired by the $q^2$ behavior for heavy-heavy
meson transitions [5].
\vskip 0.3cm
\centerline{\small Table I. $B\to\K$ form factors evaluated at $q^2=0$ and
$m_\psi^2$ in various form-factor models.}
\begin{center}
\begin{tabular}{|c||c|c|c||c|c|} \hline
 & (i) & (ii) & (iii) & BSWI & BSWII \\ \hline
$F_1^{BK}(0),~F_1^{BK}(m_\psi^2)$ & 0.56~~0.84 & 0.56~~0.84 & 0.47~~0.70 &
0.38~~0.56 & 0.38~~0.83 \\ \hline
$V^{BK^*}(0),~V^{BK^*}(m_\psi^2)$ & 0.70~~1.04 & 0.33~~0.72 & 0.70~~1.04 &
0.37~~0.55 & 0.37~~0.81 \\ \hline
$A_1^{BK^*}(0),~A_1^{BK^*}(m_\psi^2)$ & 0.55~~0.55 & 0.29~~0.41 & 0.29~~0.41 &
0.33~~0.46 & 0.33~~0.46  \\ \hline
$A_2^{BK^*}(0),~A_2^{BK^*}(m_\psi^2)$ & 0.26~~0.36 & 0.19~~0.36 & 0.31~~0.43 &
0.33~~0.46 &0.33~~0.64  \\ \hline
\end{tabular}
\end{center}

    With the $B\to\K$ form factors given in Table I, it is straightforward
to compute the quantities $R$ and $\Gamma_L/\Gamma$ (see [7,8] for their
kinematic expressions). The results are tabulated in Table II. We see that
the heavy-flavor-symmetry approach for heavy-light form factors with type
(ii) of $q^2$ dependence
gives a satisfactory agreement with the CLEO measurement of $R$ and CDF data
of $\Gamma_L/\Gamma$. However, if the recent observation of a large fraction
($\geq 70\%$) of the longitudinal polarization in $B\to\psi K^*$ by ARGUS [9]
and CLEO [1] is confirmed in the future, then case (ii) will be ruled out
either. Note that the factorization hypothesis leads to a theoretical
upper bound 0.83 for $\Gamma_L/
\Gamma$ (see e.g. Ref.[7]).  This means that if the measured $\Gamma_L/\Gamma$
is larger than 0.83, one can conclude that the factorization approach fails
irrespective of the choice of form factors. Hence, a
refined measurement of the fraction of longitudinal polarization in $B\to\psi
K^*$ is urgently called for.
\vskip 0.3 cm
\centerline{\small Table II. Predictions for $R$ and $\Gamma_L/\Gamma$
in various form-factor models.}
\begin{center}
\begin{tabular}{|c||c|c|c||c|c||c|} \hline
 & (i) & (ii) & (iii) & BSWI & BSWII & Experiment \\ \hline
$R$ & 4.00 & 1.84 & 2.83 & 4.23 & 1.61 & $1.71\pm 0.34$ [1] \\ \hline
${\Gamma_L\over\Gamma}$ & 0.61 & 0.56 & 0.41 & 0.57 & 0.35 &
$\cases{0.97\pm 0.22 &ARGUS [9] \cr
0.80\pm 0.09 &CLEO [1]\cr 0.66^{+0.13}_{-0.14}&CDF [10]\cr }$ \\ \hline
\end{tabular}
\end{center}

    Before proceeding further, we would like to give a brief summary on
how many parameters are being fitted in this procedure. To describe
$B\to\psi K(K^*)$ decays we need six form factors: $F_{0,1}^{BK},~A_{0,1,2}
^{BK^*}$ and $V^{BK^*}$ at $q^2=m_\psi^2$. They are related to $D\to K^{(*)}$
form factors at the same heavy quark velocity but near zero
recoil by model-independent heavy flavor symmetry. Using the experimental
input on $D\to K^{(*)}$ form factors at $q^2=0$, we are able to determine
$B\to K^{(*)}$ form factors at any finite $q^2$ provided that their $q^2$
dependence is known. Since $F_1$ ($A_2$) has one more pole factor than $F_0$
($A_0$ and $A_1$) and many model calculations indicate that $F_1$ is of
monopole form, it follows that only two of the form factors, say $A_2$ and
$V$, are needed to fit data to determine their $q^2$ behavior. In short,
the $q^2$ extrapolation of $F_1$ is fixed by models, while $A_2$ and $V$
by data.

     We next turn to the determination of the parameter $a_2$. Using $V_{cb}
=0.040$ [22], $\tau(B^-)=1.54\times 10^{-12}s$ [21], $f_\psi=395$ MeV
extracted from the measured width of $\psi\to e^+ e^-$ [21], and
assuming factorization, we find
\be
{\cal B}(B^-\to\psi K^-) &=& 2.85\times 10^{-2}\left|a_2F_1^{BK}(m^2_\psi)
\right|^2=\,1.99\times 10^{-2}\left|a_2\right|^2,   \non \\
{\cal B}(B^-\to\psi K^{*-}) &=& 0.221\left|a_2A_1^{BK^*}(m^2_\psi)
\right|^2=\,3.67\times 10^{-2}\left|a_2\right|^2,
\en
where we have applied type-(ii) form factors given in Table I. From the CLEO
data [1]
\be
{\cal B}(B^-\to\psi K^-) &=& (0.110\pm 0.017)\%,~~~{\cal B}(B^0\to\psi K^0)
=\, (0.075\pm 0.025)\%,   \non \\
{\cal B}(B^-\to\psi K^{*-}) &=& (0.178\pm 0.056)\%,
{}~~~{\cal B}(B^0\to\psi K^{*0}) =\, (0.169\pm 0.036)\%,
\en
the respective $|a_2|$ is found to be $0.235\pm 0.018,~0.192
\pm 0.032,~0.220\pm 0.035$, $0.212\pm 0.023$, where use of $\tau(B^0)=
1.50\times 10^{-12}s$ [21] has been made. Therefore, the combined value is
\be
\left|a_2(B\to\psi\K)\right|=\,0.221\pm 0.012\,.
\en
\vskip 0.3 cm
  {\it\bf 3. Moral from $B\to K^*\gamma$ decays}~~~We see in the previous
section that the approach of heavy flavor symmetry for heavy-light form
factors with
type (ii) of $q^2$ dependence is favored by the $B\to\psi\K$ data. We shall
see in this section that this method for heavy-light form factors also
gives an excellent description of $B\to K^*\gamma$ decay, from which useful
information on the form factors $V(0)$ and $A_1(0)$ can be extracted.

    In the standard model, the weak radiative decay $B\to K^*\gamma$ is
dominated by the short-distance penguin transition $b\to s\gamma$. The
transition amplitude for $b\to s\gamma$ reads [23]
\be
A(b\to s\gamma)=\,i{G_F\over\sqrt{2}}\,{e\over 8\pi^2}F_2(x_t)V_{tb}V_{ts}^*
\ep^\mu k^\nu\bar{s}\sigma_{\mu\nu}[m_b(1+\gamma_5)+m_s(1-\gamma_5)]b,
\en
where $F_2$ is a smooth function of $x_t\equiv m_t^2/M_W^2$ [23];
$F_2\cong 0.65$
for $m_t=174$ GeV and $\Lambda_{\rm QCD}=200$ MeV. In the static limit of the
heavy $b$ quark, we may use the equation of motion $\gamma_0 b=b$ to derive
the relation [12]
\be
\la K^*|\bar{s}i\sigma_{0i}(1\pm\gamma_5)b|B\ra =\,\la K^*|\bar{s}\gamma_i
(1\mp\gamma_5)b|B\ra.
\en
As a result, the form factors $f_1$ and $f_2$ in the matrix elements of the
tensor current can be related to the form factors $A_1$ and $V$ at the same
$q^2$. At $q^2=0$ we have (see e.g. Ref.[24])
\be
f_1^{BK^*}(0)=\,-2f_2^{BK^*}(0)=\,-\left({m_B-m_{K^*}\over m_B}V^{BK^*}(0)+
{m_B+m_{K^*}\over m_B}A_1^{BK^*}(0)\right).
\en
The decay rate is then given by
\be
\Gamma(B\to K^*\gamma)=\,{1\over 32\pi}\left({m_B^2-m^2_{K^*}\over m_B}
\right)^3\left|{G_F\over\sqrt{2}}{e\over 8\pi^2}F_2V_{tb}V_{ts}^*m_b
\right|^2\left(\left|f_1^{BK^*}(0)\right|^2+4\left|f_2^{BK^*}(0)\right|^2
\right).
\en
Hence, to the leading order in $1/m_b$ expansion
\be
{\cal B}(B\to K^*\gamma)=\,1.32\times 10^{-4}\left|A_1^{BK^*}(0)\right|^2
\left(1+0.711\,V^{BK^*}(0)/A_1^{BK^*}(0)\right)^2+{\cal O}(1/m_b^2),
\en
where use has been made of the relation $V_{tb}V_{ts}^*\cong -V_{cb}
V_{cs}^*$.

      The prediction for the branching ratio of $B\to K^*\gamma$ in the
heavy-flavor-symmetry approach for form factors with various types of $q^2$
extrapolation is exhibited in Table III. The agreement between case (ii) and
the CLEO experiment [25] is excellent. The measured branching ratio for
$B\to K^*\gamma$ together with its theoretical prediction (20) thus provides
useful constraints on the form factors $V^{BK^*}(0)$ and $A_1^{BK^*}(0)$.

\vskip 0.35cm
\centerline{\small Table III. Predictions for the branching ratio of
$B\to K^*\gamma$ in various form-factor models.}
\begin{center}
\begin{tabular}{|c||c|c|c||c|} \hline
 & (i) & (ii) & (iii) & Experiment [25] \\ \hline
${\cal B}(B\to K^*\gamma)$ & $1.5\times 10^{-4}$ & $3.6\times 10^{-5}$ &
$8.5\times 10^{-5}$  & $(4.5\pm 1.5\pm 0.9)\times 10^{-5}$ \\ \hline
\end{tabular}
\end{center}

\vskip 0.3 cm
  {\it\bf 4. Analyses of $B\to D(D^*)\pi(\rho)$ decays}~~~We learn from
previous two
sections that $B\to\psi\K$ and $B\to K^*\gamma$ decays are satisfactorily
described by the heavy-flavor-symmetry approach for heavy-light form factors
provided that $F_0$ behaves as a constant, $F_1,~A_0,~A_1$ have a monopole
behavior, and the $q^2$ dependence of $A_2$ and $V$ is of the dipole form.
In this section we will apply this method to $B\to D^{(*)}\pi(\rho)$
decays to extract $a_1$ and $a_2$.
For recent analyses of $B\to\D\pi(\rho)$ decays, see Ref.[26].

     As stressed in passing, there exist two complications for evaluating
the $\bar{B}^0\to D^{(*)+}\pi^-(\rho^-)$ decay amplitudes. First, the
$W$-exchange diagram contributes to $\bar{B}^0$ decay,
but it is difficult to estimate its size.
Second, there are final-state interactions (FSI) since the decay amplitudes
involve isospin 1/2 and 3/2 channels. In what follows we
will first analyze the data without considering these two effects, and then
come back to them later on. Based on factorization, we obtain
\be
{\cal B}(\bar{B}^0\to D^+\pi^-) &=& 0.80\times 10^{-2}\left|a_1F_0^{BD}
(m_\pi^2)\right|^2,  \non \\
{\cal B}(\bar{B}^0\to D^+\rho^-) &=& 1.93\times 10^{-2}\left|a_1F_1^{BD}
(m_\rho^2)\right|^2,  \non \\
{\cal B}(\bar{B}^0\to D^{*+}\pi^-) &=& 0.75\times 10^{-2}\left|a_1A_0^{BD^*}
(m_\pi^2)\right|^2,   \\
{\cal B}(\bar{B}^0\to D^{*+}\rho^-) &=& 0.11\times 10^{-2}\left|a_1A_1^{BD^*}
(m_\rho^2)\right|^2H,  \non
\en
and
\be
R_1\equiv{{\cal B}(B^-\to D^0\pi^-)\over{\cal B}(\bar{B}^0\to D^+\pi^-)} &=&
\r\left(1+{m_B^2-m_\pi^2\over m_B^2-m_D^2}{f_D\over f_\pi}{F_0^{B\pi}(m_D^2)
\over F_0^{BD}(m_\pi^2)}{a_2\over a_1}\right)^2,   \non \\
R_2\equiv{{\cal B}(B^-\to D^0\rho^-)\over{\cal B}(\bar{B}^0\to D^+\rho^-)} &=&
\r\left(1+{f_D\over f_\rho}{A_0^{B\rho}(m_D^2)
\over F_1^{BD}(m_\rho^2)}{a_2\over a_1}\right)^2,   \non \\
R_3\equiv{{\cal B}(B^-\to D^{*0}\pi^-)\over{\cal B}(\bar{B}^0\to D^{*+}\pi^-)}
&=& \r\left(1+{f_{D^*}\over f_\pi}{F_1^{B\pi}(m_{D^*}^2)
\over A_0^{BD^*}(m_\pi^2)}{a_2\over a_1}\right)^2,    \\
R_4\equiv{{\cal B}(B^-\to D^{*0}\rho^-)\over{\cal B}(\bar{B}^0\to D^{*+}
\rho^-)} &=& \r\left(1+2\eta{H_1\over H}+\eta^2{H_2\over H}\right),   \non
\en
with
\be
\eta &=& {m_{D^*}(m_B+m_\rho)\over m_\rho(m_B+m_{D^*})}\,{f_{D^*}\over f_\rho}
\,{A_1^{B\rho}(m^2_{D^*})\over A_1^{BD^*}(m^2_\rho)}\,{a_2\over a_1},  \non \\
H &=& (a-bx)^2+2(1+c^2y^2),   \non \\
H_1 &=& (a-bx)(a-b'x')+2(1+cc'yy'),   \\
H_2 &=& (a-b'x')^2+2(1+c'^2y'^2),   \non
\en
and
\be
a &=& {m_B^2-m_{D^*}^2-m^2_\rho\over 2m_{D^*}m_\rho},~~~b=\,{2m_B^2p_c^2\over
m_{D^*}m_\rho(m_B+m_{D^*})^2},~~~c=\,{2m_Bp_c\over (m_B+m_{D^*})^2},  \non \\
x &=& {A_2^{BD^*}(m_\rho^2)\over A_1^{BD^*}(m_\rho^2)},~~~~~y =\, {V^{BD^*}
(m_\rho^2)\over A_1^{BD^*}(m_\rho^2)},
\en
where $p_c$ is the c.m. momentum, and $b',~c',~x',~y'$ are obtained from
$b,~c,~x,~y$ respectively with the replacement $D^*\leftrightarrow \rho$; for
instance, $x'= A_2^{B\rho}(m_{D^*}^2)/ A_1^{B\rho}(m_{D^*}^2)$.

    The heavy-heavy form factors e.g. $F_0^{BD},~A_1^{BD^*}$ appearing in
(21-22) can be related to a universal Isgur-Wise function $\xi(v\cdot v')$
via heavy quark symmetry [5]. We will use the $B\to D^{(*)}$ form factors
evaluated in Ref.[5], which include $1/m_Q$ corrections.
As for the heavy-light form factor $F_1^{B\pi}$, we get $F_{0,1}^{B\pi}(0)
=0.48$ using $F_{0,1}^{D\pi}(0)\approx 0.83$ derived from $D^+\to\pi^+\pi^0$
decay [27].
\footnote{The prediction $F_{0,1}^{B\pi}(0)=0.48$ is close to the value of
0.53 obtained in the framework of chiral perturbation theory that incorporates
chiral and heavy quark symmetries [19]. Note that $F_1^{D\pi}(0)>F_1^{DK}(0)$,
whereas $F_1^{B\pi}(0)<F_1^{BK}(0)$.}
In the absence of experimental input for $D\to\rho$ form factors, we will
assume SU(3) flavor symmetry for $D\to\rho$ and $B\to\rho$ form factors at
$q^2=0$, namely $A_{1,2}^{B(D)\rho}(0)\approx A_{1,2}^{B(D)K^*}(0),
{}~V^{B(D)\rho}(0)\approx V^{B(D)K^*}(0)$.
\footnote{We have checked explicitly that, assuming $A_{1,2}^{D\rho}(0)
=A_{1,2}^{DK^*}(0)$ and $V^{D\rho}(0)=V^{DK^*}(0)$, SU(3) flavor symmetry for
$B\to\rho$ and $B\to K^*$ form factors at finite $q^2$ obtained using Eq.(6)
with type-(ii) $q^2$ dependence is better than $10\%$.}
As for the decay constants, we use $f_\pi=132$ MeV, $f_\rho=216$ MeV,
$f_D=208$ MeV [28], and
$f_{D^*}/f_D\approx 1.28$ [29]. Numerical results are summarized in Table IV,
where we have used $\tau(B^-)/\tau(B^0)=1.027$ from PDG [21].
A fit to the measured branching ratios of $\bar{B}^0\to D^{(*)}\pi(\rho)$
determines the parameter $a_1$, while the ratio $a_2/a_1$ listed in Table IV
is extracted either from the measured ratios $R_1,\cdots,R_4$ or from $B^-\to
D^{(*)}\pi(\rho)$ decays together with the $a_1$ obtained in the corresponding
$\bar{B}^0$ decays.

   The combined value of $a_1$ obtained from Table IV is
\be
a_1(B\to D^{(*)}\pi(\rho))=\,1.012\pm 0.057\,.
\en
We see that our values of $a_2/a_1$ are improved substantially over the
previous results obtained using the BSWII model [see Eq.(3)]. We also note
that the magnitude of $a_2/a_1$ determined in this manner is in general
quite stable except for $B^-\to D^{*0}\rho^-$ or $R_4$.
Therefore, a refined measurement of this decay mode is greatly
welcome. Excluding the data from $B^-\to D^{*0}\rho^-$, the combined value of
$a_2/a_1$ extracted from the remaining $B^-$ decay modes is found to be
\be
{a_2\over a_1}(B\to D^{(*)}\pi(\rho))=\,0.224\pm 0.058\,.
\en
Note that the corresponding combined value of $a_2/a_1$ obtained in the
BSWII model [see Eq.(3)] is $0.33\pm 0.08$.
Combining (26) with (25) leads to
\be
a_2(B\to D^{(*)}\pi(\rho))=\,0.226\pm 0.060\,.
\en
The main theoretical uncertainty comes from the form
factors $A_{1,2}^{B\rho}$ and $V^{B\rho}$, for which we lack
experimental input on $D\to\rho$ form factors. At this point,
we wish to emphasize again that, in contrast to Ref.[1], our value for
$a_2/a_1$ is not obtained by a least squares fit to the data.
Recall that a global fit of the BSWII model to the data yields $a_2/a_1=0.23
\pm 0.11$ [1]. Our result (27) thus improves the previous error analysis by a
factor of two.
We also remark that the fraction of longitudinal polarization in $\bar{B}^0\to
D^{*+}\rho^-$ is measured to be $0.93\pm 0.05\pm 0.05$ [1].
Unlike the $B\to\psi K^*$ case, this relative amount of longitudinal
polarization is easily accounted for by theory, which is predicted to be
$88\%$.

\vskip 0.35cm
\centerline{\small Table IV. Extraction of $a_1$ and $a_2/a_1$ from various
$B\to D(D^*)\pi(\rho)$ decays by}
\centerline{\small comparing the theoretical prediction for branching ratios
with experiment.}
\begin{center}
\begin{tabular}{|c||c|c||c|c|} \hline
 & ${\cal B}(\%)_{\rm theory}$ & ${\cal B}(\%)_{\rm expt}$ [1] & $a_1$ &
$a_2/a_1$  \\ \hline
$\bar{B}^0\to D^+\pi^-$ & $0.270a_1^2$ & $0.29\pm 0.07$ & $1.04\pm 0.13$ & -
\\ \hline
$\bar{B}^0\to D^+\rho^-$ & $0.673a_1^2$ & $0.81\pm 0.21$ & $1.10\pm 0.14$ & -
\\ \hline
$\bar{B}^0\to D^{*+}\pi^-$ & $0.260a_1^2$ & $0.26\pm 0.05$ & $1.00\pm 0.10$
& -  \\ \hline
$\bar{B}^0\to D^{*+}\rho^-$ & $0.806a_1^2$ & $0.74\pm 0.17$ & $0.96\pm 0.11$
& - \\ \hline
$B^-\to D^0\pi^-$ & $0.277a_1^2(1+1.497a_2/a_1)^2$ & $0.55\pm 0.07$
& $1.04\pm 0.13$ & $0.24\pm 0.10$  \\ \hline
$B^-\to D^{0}\rho^-$ & $0.690a_1^2(1+1.131a_2/a_1)^2$ & $1.35\pm 0.19$
& $1.10\pm 0.14$ & $0.24\pm 0.14$  \\ \hline
$B^-\to D^{*0}\pi^-$ & $0.267a_1^2(1+1.918a_2/a_1)^2$ & $0.52\pm 0.10$
& $1.00\pm 0.10$ & $0.21\pm 0.08$  \\ \hline
$B^-\to D^{*0}\rho^-$ & $ 0.827a_1^2(1+1.446a_2/a_1)^2$ & $1.68\pm 0.35$
& $0.96\pm 0.11$ & $0.34\pm 0.13$  \\ \hline
\end{tabular}
\end{center}

    Thus far we have neglected FSI and nonspectator effects. Experimentally,
FSI can be tested by measuring the decay rates
of $B^-\to D^{(*)0}\pi^-(\rho^-)$, $\bar{B}^0\to D^{(*)+}\pi^-(\rho^-),~D^{(*)
0}\pi^0(\rho^0)$ to deduce the isospin amplitudes $A_{1/2}$, $A_{3/2}$
and the phase-shift difference $(\delta_{1/2}-\delta_{3/2})$. Moreover,
in the absence of $W$ exchange, $a_2/a_1$ can be determined from $A_{3/2}/
A_{1/2}$; for example, in $B\to D\pi$ decay [30]
\be
{m_B^2-m_\pi^2\over m_B^2-m_D^2}\,{f_D\over f_\pi}\,{F_0^{B\pi}(m_D^2)\over
F_0^{BD}(m_\pi^2)}\,{a_2\over a_1}=1.497{a_2\over a_1}
=\,2{A_{3/2}/A_{1/2}-{1\over\sqrt{2}}\over A_{3/2}/A_{1/2}+\sqrt{2}}.
\en
Unfortunately, an observation of $\bar{B}^0\to D^0\pi^0$ is still not
available yet.

    The presence of $W$ exhange will affect the determination of $a_1$ from
$\bar{B}^0$ decays. Theoretically, the $W$-exchange amplitude receives its main
contribution from nonperturbative color octet currents [4], which is difficult
to estimate. Though both nonspectator and FSI effects are known to be
important in charm decays, it is generally believed that they do not play a
significant role in bottom decay as the decay particles are moving fast, not
allowing adequate time for FSI.
\vskip 0.3 cm
   {\it\bf 5. Discussion and conclusion}~~~ We see from (15) and (27) that
the magnitude of $a_2$ determined from $B\to\D\pi(\rho)$ decays  agrees
well with that extracted from $B\to\psi\K$. Then, can we conclude that $a_2$
extracted in the latter decay is positive? Recall that, as we have argued
before, nonperturbative effects must be in such a way that $|r_2(B\to\psi\K)|~
{\Large^{>}_{\sim}}~|r_2(B\to\D\pi)|$ [4]. Since $a_2(B\to\D\pi)=(c_2+
c_1/N_c)+{1\over 2}
r_2c_1$ is positive and $c_2+c_1/N_c=0.15\sim 0.20$ at $\mu=m_b$ beyond
the leading logarithmic approximation [31],
\footnote{To the leading logarithmic approximation, $c_1(m_b)\sim 1.11,
{}~c_2(m_b)\sim -0.26$ and hence $c_2+c_1/N_c\sim 0.11$.}
it is clear that $r_2(B\to\D\pi)=0.05\sim 0.14$ is positive. Now there are
two possibilities for $a_2$ in $B\to\psi\K$: (i) $r_2>0$; this implies a
positive $a_2$ and that $a_2(B\to\psi\K)~{\Large ^{>}_{\sim}}~a_2(B\to\D\pi)
$, and (ii) $r_2<0$; this together with (15) indicates a negative $a_2$ and
$r_2(B\to\psi\K)=-(0.72\sim 0.80)$. We consider the case (ii) very unlikely
since the magnitude of $r_2$ in $B\to\psi\K$ should not deviate too much from
that in $B\to\D\pi$. Therefore, contrary to the previous publication [4],
we believe that $a_2(B\to\psi\K)$ ought to be positive. In fact, it is not
difficult to achieve the relation
$a_2(B\to\psi\K)~{\Large ^{>}_{\sim}}~a_2(B\to\D\pi)$. Note
that our extraction of $a_{1,2}$ from $B\to\D\pi(\rho)$ so far is based on the
assumption that FSI and $W$ exchange are negligible. It is likely that
the inclusion of these two effects will reduce the present estimate of
$a_2(B\to\D\pi(\rho))$. In view of this, a measurement of $\bar{B}^0\to
D^{(*)0}\pi^0(\rho^0)$ is urgently needed. Finally, the question of why $r_2$
is positive in exclusive $B$ decays whereas it is negative in $D$
decays remains an enigma. This will be a great challenge to both lattice
and QCD-sum-rule practitioners. Nevertheless, the small magnitude of $r_2$
in $B\to\D\pi(\rho)$ compared to that in exclusive
two-body $D$ decays ($|r_2|=(0.67\sim 1.3)$) is consistent with our
expectation that nonfactorizable soft-gluon effects are much less significant
in the former. This also explains why the $1/N_c$ approach, which is
empirically known to be operative in charm decays, fails in
$B\to D^{(*)}\pi(\rho)$ and $B\to \psi K^{(*)}$ decays.

    To conclude, we have shown that, based on factorization, the
heavy-flavor-symmetry approach for heavy-light form factors in conjunction
with the type-(ii) $q^2$ dependence provides a satisfactory description of the
CLEO data on ${\cal B}(B\to \psi K^*)/{\cal B}(B\to\psi K)$
and the CDF data on the fraction of the longitudinal polarization in $B\to\psi
K^*$ decays. However, if the measured $\Gamma_L/\Gamma$ is larger than 0.70,
then our form factors are also ruled out. Furthermore, we will conclude
that the factorization approach fails irrespective the choice of form factors
if the relative amount of the transverse
polarization is found to be less than 17\%.
Therefore, a refined measurement of $\Gamma_L/\Gamma$ in $B\to\psi K^*$
is urgently needed in order to test form-factor models and the factorization
hypothesis. Armed with the above method, we
have extracted the parameters $a_1$ and $a_2$ from $B\to\psi\K$ and $B\to
\D\pi(\rho)$ decays. Our result $a_2/a_1=0.22\pm 0.06$ improves the previous
error analysis by a factor of two. Finally, we have argued that the sign of
$a_2(B\to\psi K^{(*)})$ should be positive and we have discussed its
important implications.

\pagebreak
\centerline{\bf ACKNOWLEDGMENT}

    This work was supported in part by the National Science Council of ROC
under Contract No. NSC84-2112-M-001-014.

\vskip 0.8 cm
\centerline{\bf REFERENCES}
\vskip 0.3 cm
\begin{enumerate}

\item CLEO Collaboration, M.S. Alam {\it et al.,} \pr {\bf D50}, 43 (1994).

\item A.J. Buras, J.-M. G\'erard, and R. R\"uckl, \np {\bf B268}, 16 (1986).

\item M. Bauer, B. Stech, and M. Wirbel, \zp {\bf C34}, 103 (1987).

\item H.Y. Cheng, \pl {\bf B335}, 428 (1994).

\item M. Neubert, V. Rieckert, B. Stech, and Q.P. Xu, in {\it Heavy Flavors},
edited by A.J. Buras and H. Lindner (World Scientific, Singapore, 1992).

\item N. Deshpande, M. Gronau, and D. Sutherland, \pl {\bf 90B}, 431 (1980).

\item M. Gourdin, A.N. Kamal, and X.Y. Pham, PAR/LPTHE/94-19 (1994).

\item R. Aleksan, A. Le Yaouanc, L. Oliver, O. P\`ene, and J.-C. Raynal,
DAPNIA/SPP/94-24, LPTHE-Orsay 94/15 (1994).

\item ARGUS Collabotation, DESY 94-139 (1994).

\item CDF Collaboration, FERMILAB Conf-94/127-E (1994).

\item C.E. Carlson and J. Milana, WM-94-110 (1994).

\item N. Isgur and M. Wise, \pr {\bf D42}, 2388 (1990).

\item M. Wirbel, B. Stech, and M. Bauer, \zp {\bf C29}, 637 (1985).

\item Q.P. Xu, \pl {\bf B306}, 363 (1993).

\item C.A. Dominguez and N. Paver, \zp {\bf C41}, 217 (1988); A.A.
Ovchinnikov, {\sl Sov. J. Nucl. Phys.} {\bf 50}, 519 (1989); \pl {\bf B229},
127 (1989); V.L. Chernyak and I.R. Zhitnitski, \np {\bf B345}, 137 (1990);
S. Narison, \pl {\bf B283}, 384 (1992); V.M. Belyaev, A. Khodjamirian, and
R. R\"uckl, \zp {\bf C60}, 349 (1993); P. Colangelo, BARI-TH/93-152 (1993).

\item P. Ball, V.M. Braun, and H.G. Dosch, \pl {\bf B273}, 316 (1991); \pr
{\bf D44}, 3567 (1991); P. Ball, \pr {\bf D48}, 3190 (1993).

\item P. Colangelo and P. Santorelli, \pl {\bf B327}, 123 (1994).

\item K.C. Yang and W-Y.P. Hwang, NUTHU-94-17 (1994).

\item H.Y. Cheng, {\sl Chin. J. Phys.} {\bf 32}, 425 (1994).

\item M. Witherell, in {\it Proceedings of the XVI International Symposium
on Lepton-Photon Interactions}, Ithaca, 10-15 August 1993, eds. P. Drell
and D. Rubin (AIP, New York, 1994).

\item Particle Data Group, \pr {\bf D50}, 1173 (1994).

\item M. Neubert, \pl {\bf B338}, 84 (1994).

\item M. Misiak, \pl {\bf B269}, 161 (1991); \np {\bf B393}, 23 (1993); G.
Cella, G. Curci, G. Ricciardi, and A. Vicer\'e, \pl {\bf B248}, 181 (1990);
{\sl ibid.} {\bf B325}, 227 (1994); B. Grinstein, R. Springer, and M.B. Wise,
\np {\bf B339}, 269 (1990); R. Grigjanis, P.J. O'Donnell, M.
Sutherland, and H. Navelet, \pl {\bf B213}, 355 (1988); {\sl ibid.}
{\bf B286}, 413(E) (1992); {\sl Phys. Rep.} {\bf 228}, 93 (1993).

\item P.J. O'Donnell and H.K.K. Tung, \pr {\bf D48}, 2145 (1993).

\item CLEO Collaboration, R. Ammar {\it et al.,} {\sl Phys. Rev. Lett.}
{\bf 71}, 674 (1993).

\item A. Deandrea, N. Di Bartolomeo, R. Gatto, and G. Nardulli, \pl {\bf
B318}, 549 (1993); K. Honscheid, K.R. Schubert, and R. Waldi, \zp {\bf C63},
117 (1994); S. Resag and M. Beyer, \zp {\bf C63}, 121 (1994);
M. Gourdin, A.N. Kamal, Y.Y. Keum, and X.Y. Pham, \pl {\bf B333}, 507 (1994);
A.N. Kamal and T.N. Pham, \pr {\bf D50}, 395 (1994).

\item L.L. Chau and H.Y. Cheng, \pl {\bf B333}, 514 (1994).

\item C.W. Bernard, J.N. Labrenz, and A. Soni, \pr {\bf D49}, 2536 (1994).

\item A. Abada {\it et al.,} \np {\bf B376}, 172 (1992); M. Neubert, \pr
{\bf D46}, 1076 (1992).

\item H. Yamamoto, HUTP-94/A006 (1994).

\item A.J. Buras, MPI-PhT/94-60 (1994).

\end{enumerate}

\end{document}